\documentclass[superscriptaddress,aps,pra,twocolumn,showpacs,floatfix]{revtex4-2}
\usepackage[utf8]{inputenc}
\usepackage{graphicx,amsmath,amsfonts,amssymb}
\usepackage{color}
\usepackage[colorlinks=true, allcolors={blue}]{hyperref}
\usepackage{graphicx}
\usepackage{epstopdf}
\usepackage{float}
\usepackage{placeins}
\newcommand{\orcid}[1]{\href{https://orcid.org/#1}{\includegraphics[width=7pt]{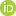}}}

\begin{document}
\preprint{APS/123-QED}

\title{Enhanced frequency and temperature estimation by a $\mathcal{PT}$-symmetric
quantum oscillator}

\author{Jonas F. G. Santos \orcid{0000-0001-9377-6526}}
\email{jonassantos@ufgd.edu.br}
\affiliation{Faculdade de Ci\^{e}ncias Exatas e Tecnologia, Universidade Federal da Grande Dourados, Caixa Postal 364, CEP 79804-970, Dourados, MS, Brazil}

\begin{abstract}
Quantum metrology employs quantum properties to enhance the precision
of physical parameters, in order to characterize quantum states as
well as channels. Frequency and temperature estimations are of fundamental
importance for these tasks and have been considerably treated in quantum
sensing strategies. From the set of quantum features that can be exploited
in quantum metrology, those related to non-Hermitian systems have
received much attention in recent times. Here, we consider a paradigmatic
non-Hermitian system, the quantum Swanson oscillator, as a probe system,
in order to investigate how the non-Hermitian contribution affects
the frequency and temperature estimation. We use the quantum Fisher
information to compute the main results. Furthermore, to perform a
fair comparison, we define a gain function which is the ratio between
the quantum Fisher information with the non-Hermitian contribution
and the quantum Fisher information when this term becomes Hermitian. In addition, the quantum Fisher
information is discussed in terms of the energetic cost to include
the non-Hermitian contribution. Given that we obtain the Hermitian
counterpart of the Hamiltonian by applying the Dyson map, we also
study the estimation of the parameters characterizing this mapping.
Our results indicate that the non-Hermitian contribution in the Swanson
quantum oscillator can contribute to enhance the frequency and temperature
estimation.
\end{abstract}

\maketitle

\section{Introduction}

The rapid development of quantum technologies in the last years has
revealed that for a given task a quantum system can perform better
than the corresponding classical one \citep{Auff=0000E8ves2022,Mukhopadhyay2025}.
Parameter estimation using quantum systems has received considerable
attention in this scenario. Quantum parameter estimation employs intrinsic
quantum features, such as coherence or quantum correlations to obtain
a precision gain beyond the standard quantum limit \citep{Wiseman2010,Degen2017}.
The current state-of-the-art in quantum parameter estimation includes
a plethora of different systems, such as trapped ion systems \citep{Marciniak2022,Ding2021},
superconducting qubits \citep{Wolski2020,Danilin2024}, and single
photons \citep{Motes2015,Pirandola2018}. From the point of view of
quantum resources to perform parameter estimation, protocols have
exploited quantum coherence \citep{Santos2024}, quantum correlations
\citep{Lee2020,Porto2025}, as well as squeezing for bosonic or spin
systems \citep{Gross2012,Zhang2014,Mukhopadhyay2025-1}. Furthermore,
systems undergoing quantum phase transitions are also strong candidates to achieve high performance in estimating parameters. \citep{Liu2016,Garbe2020}. Moreover,
any realistic implementation of quantum sensors has to take into account
the energetic balance in quantum systems, and recently the role played by thermodynamics and
irreversibility has been investigated in quantum parameter estimation
protocols \citep{Chu2022}.

Intrinsic quantum features are not restricted to coherence or quantum
correlations. In recent times, proposals for quantum parameter estimation
protocols have exploited the properties of non-Hermitian physics \citep{Ding2021,Wang2024}.
Although the standard quantum mechanics (SQM) assumes that the Hermiticity
of a Hamiltonian operator guarantees a real set of eigenvalues and
probability conservation, Bender and Boettcher \citep{Bender1998,Bender2015}
proposed that a non-Hermitian Hamiltonian also possesses real spectra
provided it fulfills the conditions of invariance by spatial reflection
(parity $\mathcal{P}$) and time reversal ($\mathcal{T}$). Hamiltonian
operators satisfying these conditions are called $\mathcal{PT}$-
symmetric Hamiltonians and have been investigated in different branches
of physics, such as fluctuation relations \citep{Deffner2015,Zeng2017},
photonics systems \citep{Regensburger2002,Ruter2010} and time-dependent
Hamiltonians \citep{Faria2007,Ponte2019}. For quantum systems described
by $\mathcal{PT}$- symmetric Hamiltonians, the probability conservation
is preserved by including a metric operator \citep{Mostafazadeh2004}.
This allows to introduce the concept of pseudo-Hermiticity and, in
fact, given a $\mathcal{PT}$- symmetric Hamiltonian, one can obtain
the Hermitian counterpart by using the well-known Dyson map \citep{Mostafazadeh2004}.
Thus, all the relevant non-Hermitian features are inserted in the
Hermitian counterpart. 

A paradigmatic model in the context of non-Hermitian $\mathcal{PT}$-
symmetric systems is the well-known Swanson oscillator \citep{Graefe2015,Swanson2004,Midya2011,Fern=0000E1ndez2022,Sinha2024}.
It can be described by a single bosonic mode or a cavity plus a non-Hermitian
term, often given in terms of quadratic creator and annihilator operators.
The Swanson oscillator can be mapped to its Hermitian counterpart
using different Dyson maps and it is verified that in some cases both
the Swanson oscillator and its Hermitian counterpart are isospectral
\citep{Midya2011}. The classical version of the Swanson oscillator
has also been investigated, where the phase-space trajectories are
used to sign non-Hermiticity features \citep{Graefe2015}. Besides
the Swanson oscillator, Ref. \citep{Longhi2016} has also proposed
a different $\mathcal{PT}$- symmetric quantum oscillator in optical
cavity where the simulation is based on transverse light dynamics
in a ressonator with spatially-inhomogeneous gain. 

In view of the versatility of the Swanson oscillator, we here propose
to employ it as a probe system in a quantum parameter estimation protocol.
We investigate the role played by the non-Hermitian contribution (n-HC) in
the frequency and temperature estimation, two relevant quantities
that often characterize quantum states. To do so, we employ the quantum
Fisher information (QFI) as well as a gain function defined later.
For this purpose, we first use the metric operator to show the equivalence
between expectation values computed using general states engendered
by eigenstates associated with the $\mathcal{PT}$-symmetric Hamiltonian
or with its Hermitian counterpart. Then, the QFI is computed for the
relevant physical parameters of the system, and the effect due to
the  n-HC is quantified by defining a gain function,
which is the ratio between the QFI with the non-Hermitian contribution
and the QFI when this term becomes Hermitian. We also compare the
QFI for frequency and temperature estimation with an energy difference
which measures the energetic cost of including the  n-HC
in the Swanson Hamiltonian. Given that the Dyson map is used to obtain
the Hermitian counterpart, we also study the QFI for the parameter
associated with this map, which is related to the  n-HC.
Our results indicate that the  n-HC in the Swanson
Hamiltonian can be exploited to enhance the QFI for frequency and
temperature estimation. Furthermore, the results are not restricted to a specific
set of parameters in the Swanson model, although for some set of them
the enhancement is not achieved.

The present work is organized as follows. Section \ref{sec:Theoretical-framework}
establishes the grounds of quantum parameter estimation for Gaussian
states and $\mathcal{PT}$-symmetric quantum mechanics. We also discuss
the relation between observables in the Hermitian and non-Hermitian
framework of quantum mechanics for mixed states, as well as the role
played by the  n-HC in parameter estimation protocols.
A gain ratio is introduced and, additionally,  we detail our scheme to
perform estimation with a $\mathcal{PT}$-symmetric Hamiltonian and
a similarity transformation. In Section \ref{sec:Example:--symmetric-single}
we consider the quantum Swanson oscillator as the probe system and
we investigate the parameter estimation in view of the QFI, the gain
ratio, and the energetic cost of including the n-HC.
It is also observed that there exists a critical value for the parameter
associated to the n-HC such that the QFI diverges,
theoretically allowing an infinite precision. We draw the final remarks
and conclusion in \ref{sec:Conclusion}.

\section{Theoretical framework\label{sec:Theoretical-framework}}

\textbf{Quantum parameter estimation. }Consider a general parameter
$\theta$ that can be encoded in some quantum state due to some quantum
operation, with squared sensitivity denoted by $\left(\delta\theta\right)^{2}$.
In order to have an estimation of $\theta$, we collect $Q$ measurements
results $a_{i}$ of some observable $A$ and define the variance of
the deviation from the true value of $\theta$ using some estimator,
which depends solely on the measurement outcomes. The precision in
estimating $\theta$ is bounded from below by the inverse of the
quantum Fisher information (QFI)

\begin{equation}
\left(\delta\theta\right)^{2}\geq\frac{1}{Q\mathcal{I_{\theta}}},\label{bound}
\end{equation}
with $\mathcal{I}_{\theta}=\mathcal{I}_{\theta}\left[\rho\left(\theta\right)\right]$
the quantum Fisher information for a single-parameter estimation,
which is basically the optimization of the classical Fisher information
$I_{\theta}$ over all possible positive operator-valued measure (POVM),
$\mathcal{I_{\theta}}=\sup_{K_{i}}I_{\theta}$, with $K_{i}$ the
associated POVM such that $\sum_{i}K_{i}^{\dagger}K_{i}=\mathbb{I}$.
From Eq. (\ref{bound}) we note that the higher the QFI the less the possible
uncertainty in the measurement of $\theta$, and then we search for
probe states that provide a QFI as sensitive as possible to small
parameter variations. Because of this fact, the QFI can be related
to different distance quantifiers \citep{Serafinibook,Rosati2018,Fanizza2021}.
In particular, for the Bures distance between two close states $\rho_{\theta}$
and $\rho_{\theta+\epsilon}$, with $\epsilon\ll1$ and defined as

\begin{equation}
d_{B}\left(\epsilon\right)=\sqrt{2}\sqrt{1-\sqrt{\mathcal{F}\left(\rho_{\theta},\rho_{\theta+\epsilon}\right)}},\label{bures}
\end{equation}
the QFI is written as

\begin{equation}
\mathcal{I}\left(\rho_{\theta}\right)=4\left(\frac{\partial d_{B}\left(\epsilon\right)}{\partial\epsilon}\right)^{2}\rvert_{\epsilon=0}.\label{QFI01}
\end{equation}

For any single-mode Gaussian state, the fidelity $\mathcal{F}\left(\rho_{\alpha},\rho_{\beta}\right)=\left(\text{Tr}\sqrt{\sqrt{\rho_{\alpha}}\rho_{\beta}\sqrt{\rho_{\alpha}}}\right)^{2}$
depends only on the first and second statistical moments, $\langle d_{i}\rangle_{\rho}$
and $\sigma_{ij}=\langle d_{i}d_{j}+d_{j}d_{i}\rangle_{\rho}-2\langle d_{i}\rangle_{\rho}\langle d_{j}\rangle_{\rho}$,
respectively, with the vector $\vec{d}=\left(x,p\right)$. It is given
by

\begin{equation}
\mathcal{F}\left(\rho_{\alpha},\rho_{\beta}\right)=\frac{2}{\sqrt{\Delta+\delta}-\sqrt{\delta}}\exp\left[-\frac{1}{2}\Delta\vec{d}^{T}\left(\Sigma_{\alpha}+\Sigma_{\beta}\right)^{-1}\Delta\vec{d}\right],
\end{equation}
with $\Delta\equiv\det\left[\boldsymbol{\sigma}_{\alpha}+\boldsymbol{\sigma}_{\beta}\right]$,
$\delta\equiv(\text{det}\boldsymbol{\sigma}_{\alpha}-1)(\text{det}\boldsymbol{\sigma}_{\beta}-1),$and
$\Delta\vec{d}=\vec{d}_{\alpha}-\vec{d}_{\beta}.$ Expanding the fidelity
up to the second order in $\epsilon$ and using Eqs. (\ref{bures})
and (\ref{QFI01}), the QFI for a Gaussian probe can finally be expressed
as \citep{Pinel2013}

\begin{equation}
\mathcal{I}\left(\rho_{\theta}\right)=\frac{1}{2}\frac{\text{Tr}\left[\left(\boldsymbol{\sigma}_{\theta}^{-1}\boldsymbol{\sigma}_{\theta}'\right)^{2}\right]}{1+P_{\theta}^{2}}+2\frac{(P_{\theta}^{'})^{2}}{1-P_{\theta}^{4}}+\Delta\vec{d}'^{T}\boldsymbol{\sigma}_{\theta}^{-1}\Delta\vec{d}',\label{QFI}
\end{equation}
with `` $'$ '' indicating derivative with respect to $\theta$,
and $P_{\theta}=|\boldsymbol{\sigma}_{\theta}|^{-2}$ representing
the purity. Note that, eventually, the probe system can depend on
more than one parameter, and in this case we can write $\rho^{th}\left(\vec{\theta}\right)$,
with $\vec{\theta}=\left(\theta_{1},...,\theta_{J}\right)$, with
$J$ the total number of parameters characterizing the probe state.
The QFI in Eq. (\ref{QFI}) is still valid since we are estimating
one parameter at a time.

\textbf{$\mathcal{PT}$-symmetric quantum mechanics. }Any physical
observable is characterized by a set of real eigenvalues. To ensure
this, standard quantum mechanics imposes that in order to be an observable,
any operator must have a real spectra and a set of complete eigenstates.
Hermitian operators, $O=O^{\dagger}$, satisfy these two conditions.
However, this is not the only class of operators fulfilling the conditions
for being an operator. It has been shown in Ref. \citep{Bender1998}
that if an operator is simultaneously invariant under parity $\mathcal{P}$
and time reversal $\mathcal{T}$ then it also possesses real spectra
and a complete set of eigenstates, becoming a possible observable
for some physical quantity. Taking a general $N$- dimensional Hamiltonian
$\mathcal{H}\left(q_{j},p_{j}\right)$ as a toy model, with $j=1,...,N$,
the unbroken $\mathcal{PT}$- symmetry guarantees the reality of the
spectrum of $\mathcal{H}\left(q_{j},p_{j}\right)$. Mathematically,
this implies that $\left[\mathcal{H}\left(q_{j},p_{j}\right),\mathcal{PT}\right]=0$,
as well as $\mathcal{PT}|\Psi_{n}\left(t\right)\rangle=\Psi_{n}\left(t\right)\rangle$,
with $\Psi_{n}\left(t\right)\rangle$ the eigenstates of $\mathcal{H}\left(q_{j},p_{j}\right)$,
and as a consequence the Hamiltonian must be invariant under the following
set of transformations \citep{Faria2007}

\begin{align}
\mathcal{PT}q_{j}\left(\mathcal{PT}\right)^{-1} & \rightarrow-q_{j},\nonumber \\
\mathcal{PT}p_{j}\left(\mathcal{PT}\right)^{-1} & \rightarrow p_{j},\nonumber \\
\mathcal{PT}i\left(\mathcal{PT}\right)^{-1} & \rightarrow-i.\label{eq01}
\end{align}

A Hamiltonian $\mathcal{H}=\mathcal{H}\left(q_{j},p_{j}\right)$ that
is invariant under the transformation in Eq. (\ref{eq01}), i.e. $\mathcal{H}\left(q_{j},p_{j}\right)=\mathcal{H}^{\mathcal{PT}}\left(q_{j},p_{j}\right)$,
is called $\mathcal{PT}$-symmetric Hamiltonian. Furthermore, given
a $\mathcal{PT}$-symmetric Hamiltonian $\mathcal{H}$, it admits
a Hermitian partner through the similarity transformation \citep{Bender2019,Fring2016,Luiz2020,Tai2023}

\begin{equation}
H=\eta\mathcal{H}\eta^{-1},\label{eq02}
\end{equation}
where $\eta=\eta\left(q_{j},p_{j}\right)$ is the Dyson map with the
condition $\eta\eta^{-1}=\mathbb{I}$, with $\mathbb{I}$ the identity
operator. By using Eq. (\ref{eq02}) and the Hermiticity relation,
it can be proven that $\Theta\mathcal{H}=\mathcal{H}^{\dagger}\Theta$,
known as quasi-Hermiticity relation, with $\Theta=\eta^{\dagger}\eta$
the metric operator to ensure probability conservation in the non-Hermitian
quantum mechanics (NHQM) \citep{Fring2016,Luiz2020}. 

\textbf{Observable}. An essential ingredient in quantum mechanics
is the evaluation of expectation values, in which for the NHQM, the
metric operator plays a fundamental role. The similarity transformation
is equally valid to any operator, not only the Hamiltonian. This means
that given a non-Hermitian operator $\mathcal{O}$, its Hermitian
counterpart is obtained by $O=\eta\mathcal{O}\eta^{-1}$\citep{Luiz2020}.
From this, it is direct to show that expectation values for these
observables are the same, i.e.,

\begin{equation}
\langle\phi_{n}\left(t\right)|O|\phi_{n}\left(t\right)\rangle=\langle\psi_{n}\left(t\right)|\Theta\mathcal{O}|\psi_{n}\left(t\right)\rangle,\label{eq:eq03}
\end{equation}
with $|\phi_{n}\left(t\right)\rangle=\eta\psi_{n}\left(t\right)\rangle$
and, for clarity, $\left\{ |\phi_{n}\left(t\right)\rangle\right\} $(
$\left\{ |\psi_{n}\left(t\right)\rangle\right\} $) forms a basis
in the standard (non-Hermitian) quantum mechanics. This important
feature allows that, given a $\mathcal{PT}$-symmetric Hamiltonian
$\mathcal{H}$ we can simply choose a similarity transformation (Dyson
map) to obtain the Hermitian counterpart $H$ and then work out the
computation of the expectation values in the standard quantum mechanics.

Despite the relevance of the equation (\ref{eq:eq03}), it holds for
pure states of a given system. For the purpose of the present work,
we would like to extend this expression for mixed states, i.e., describing
the system state using a density matrix. For simplicity we focus on
thermal states, which is a special class of density matrix with the
general form in the SQM

\begin{equation}
\rho^{th}=\sum_{n}c_{n}|\phi_{n}\rangle\langle\phi_{n}|,\label{eq:aa}
\end{equation}
with $c_{n}$ representing a thermal distribution fulfilling $\sum_{n}c_{n}=1$.
Using $|\phi_{n}\left(t\right)\rangle=\eta\psi_{n}\left(t\right)\rangle$
it is straightforward to show (see appendix) that the relation between
$\rho^{th}$ and its non-Hermitian counterpart $\tilde{\rho}^{th}$
is given by

\begin{equation}
\rho^{th}=\eta\tilde{\rho}^{th}\eta^{\dagger},\label{eq04=00005B}
\end{equation}
with $\tilde{\rho}^{th}\equiv\sum_{n}c_{n}|\psi_{n}\rangle\langle\psi_{n}|$.
We highlight that Eq. (\ref{eq04=00005B}) together with the definition
of $\tilde{\rho}^{th}$ guarantee that the population $c_{n}$ is kept
invariant under similarity transformation. We are now in position
to obtain the relation for expectation values for mixed states. In
the appendix we show that the equality for expectation values holds for
the thermal states relation in Eq. (\ref{eq04=00005B}) and is given
by

\begin{equation}
\langle O\rangle_{\rho^{th}}=\langle\Theta^{2}\mathcal{O}\rangle_{\tilde{\rho}_{th}}.\label{eq05}
\end{equation}

The extension of previous results of expectation values of observables
from pure states to mixed states allows to consider $\mathcal{PT}$-symmetric
Hamiltonians in more general protocols, for instance, in quantum metrology
where the set of parameter to be estimated is encoded in the thermal
distribution $c_{n}$. 

\subsection*{Role played by $\mathcal{PT}$-symmetric Hamiltonians in the parameter
estimation }

Here we detail how $\mathcal{PT}$-symmetric Hamiltonians can be employed
in parameter estimation through the QFI. The first point to be highlighted
concerns the statistical moments in Eq. (\ref{QFI}). Equations (\ref{eq:eq03})
and (\ref{eq05}) have shown that the expectation values, even for
pure or mixed states, are the same evaluated in the SQM or in the
NHQM. This allows two possible ways to obtain expectation values in
Eq. (\ref{QFI}): we can calculate $\mathcal{O}$ by applying the
similarity transformation and then use the metric operator to evaluate
$\langle\Theta^{2}\mathcal{O}\rangle_{\tilde{\rho}_{th}}$; or we
can apply the similarity transformation directly on the Hamiltonian
as in Eq. (\ref{eq02}) and then construct the state $\rho_{th}$
to compute $\langle O\rangle_{\rho^{th}}$. For convenience we choose
the second route. 

The scheme to study the n-HC in the frequency
and temperature estimation is organized in the following steps:

$a)$ Given a non-Hermitian Hamiltonian $\mathcal{H}$ fulfilling
the $\mathcal{PT}$-symmetry, we apply the similarity transformation
$H=\eta\mathcal{H}\eta^{-1}$ to obtain the Hermitian counterpart
$H$ acting on the standard quantum mechanics;

$b)$ The thermal state $\rho^{th}=e^{-\beta H}/Z$, with $Z$ the
partition function, is prepared for the probe system, where the non-Hermitian
contribution is encoded in one or more parameters in $H$;

$c)$ The frequency and temperature estimation is performed by computing
the QFI $\mathcal{I}_{\omega}\left[\rho^{th}\left(\vec{\theta}\right)\right]$
and $\mathcal{I}_{\beta}\left[\rho^{th}\left(\vec{\theta}\right)\right]$,
with $\vec{\theta}$ here representing all parameters characterizing
the probe state, including the frequency and temperature, as well
as those parameters of the similarity transformation. 

The second aspect concerns the action of the similarity transformation
$\eta$. Suppose that $\eta=\eta\left(\vec{\epsilon}\right)$, i.e.,
the similarity transformation depends on the family of parameters
$\vec{\epsilon}=\left(\epsilon_{1},...,\epsilon_{M}\right)$, with
$M$ the number of parameters. In this case, there is a vector $\vec{\epsilon}_{\text{Herm}}$
which corresponds to $\eta=\mathbb{I}$ or, equivalently, to $\mathcal{H}$
to be Hermitian by nature. To have a fair comparison concerning the
use of the  n-HC of a Hamiltonian in the parameter
estimation, we define a Gain Ratio $\tau_{\vec{\epsilon}}^{\theta_{i}}$
as

\begin{equation}
\tau_{\vec{\epsilon}}^{\theta_{i}}\equiv10\log\left\{ \frac{\mathcal{I}_{\theta_{i}}\left[\rho^{th}\left(\vec{\theta}\right)\right]}{\mathcal{I}_{\theta_{i}}\left[\rho_{\vec{\epsilon}_{\text{Herm}}}^{th}\left(\vec{\theta}\right)\right]}\right\} ,\label{raito}
\end{equation}
with $\mathcal{I}_{\theta_{i}}\left[\rho^{th}\left(\vec{\theta}\right)\right]$
and $\mathcal{I}_{\theta_{i}}\left[\rho_{\vec{\epsilon}_{\text{Herm}}}^{th}\left(\vec{\theta}\right)\right]$
the QFI using a non-Hermitian $\mathcal{PT}$-symmetric Hamiltonian
$\mathcal{H}=\mathcal{H}^{\mathcal{PT}}$ as a probe and the QFI using
the Hermitian $\mathcal{H}=\mathcal{H}^{\dagger}$, by imposing $\eta\left(\vec{\epsilon}_{\text{Herm}}\right)=\mathbb{I}$,
as a probe system, respectively. The quantity $\tau_{\vec{\epsilon}}^{\theta_{i}}$
is a convenient definition that indicates when the  n-HC
provides an enhancement in the estimation of a general parameter $\theta_{i}$,
i.e., when $\tau_{\vec{\epsilon}}^{\theta_{i}}>0$. Furthermore, from
$\eta=\eta\left(\vec{\epsilon}\right)$ we can also compute the QFI
for each of these parameters. 

\section{The quantum Swanson oscillator as a probe system\label{sec:Example:--symmetric-single}}

We start with a brief review on the quantum Swanson oscillator \citep{Swanson2004,Jones2005},
which the time-independent Hamiltonian is 

\begin{equation}
\mathcal{H}_{S}=\omega a^{\dagger}a+\alpha a^{2}+\beta a^{\dagger2},\label{Swanson01}
\end{equation}
with $H_{\text{Herm}}=\omega a^{\dagger}a$, $H_{\text{NH}}=\alpha a^{2}+\beta a^{\dagger2}$
the Hermitian and non-Hermitian contribution, respectively, $a\left(a^{\dagger}\right)$
standing for the annihilation (creation) bosonic operator, $\alpha,\beta\in\mathbb{R}$
and $\hbar=1$. The condition $\alpha\neq\beta$ ensures that the
Hamiltonian is not Hermitian, $\mathcal{H}_{S}\neq\mathcal{H}_{S}^{\dagger}$,
whereas it is Hermitian for $\alpha=\beta$ and in this case has been
extensively treated in the time-dependent scenario in Ref.\citep{Onah2023}.
The time-dependent Swanson Hamiltonian has been also considered in
Ref. \citep{Fring2016-1}. Writing the quadrature operators as $q=\sqrt{\hbar/\left(2\omega\right)}\left(a^{\dagger}+a\right)$
and $p=i\sqrt{\hbar\omega/2}\left(a^{\dagger}-a\right)$, with $m\equiv1$,
and using Eq. (\ref{eq01}) it is easy to show that $\mathcal{PT}$-symmetry
implies $a\rightarrow-a$ and $a^{\dagger}\rightarrow-a^{\dagger}$,
which shows that the Swanson Hamiltonian is $\mathcal{PT}$-symmetric,
$\mathcal{H}_{S}=\mathcal{H}_{S}^{\mathcal{PT}}.$ The Swanson Hamiltonian
is pseudo-Hermitian, in the sense that it is connected to its Hermitian
counterpart by $H_{S}=\eta\mathcal{H}_{S}\eta^{-1}$. As pointed out
in Ref. \citep{Jones2005}, for $\omega>\alpha+\beta$, the energy
spectrum of $\mathcal{H}_{S}$ corresponds to that of a single quantum
harmonic oscillator with frequency $\Omega=\sqrt{\omega^{2}-4\alpha\beta}$.
In principle, the choice of $\eta$ is arbitrary, but it depends on
$\alpha$ and $\beta$ in $H_{\text{NH}}$. Here we consider, without
loss of generality, the Swanson Hamiltonian such that $\alpha\beta=\omega^{2}\epsilon^{2}$
, such that the Eq. (\ref{Swanson01}) becomes

\begin{align}
\mathcal{H}_{S} & =\omega a^{\dagger}a+\alpha a^{2}+\frac{\omega^{2}\epsilon^{2}}{\alpha}a^{\dagger2},\label{Swanson02}
\end{align}
clearly non-Hermitian but $\mathcal{PT}$-symmetric, which corresponds
to a quantum harmonic oscillator with frequency $\Omega=\omega\sqrt{1-4\epsilon^{2}}$
. By choosing the Dyson map to be $\eta=\exp\left[\frac{1}{2}\frac{\left(1-\omega^{2}\epsilon^{2}\right)}{\left(1-\omega+\omega^{2}\epsilon^{2}\right)}x^{2}\right]$
\citep{Jones2005}, the Hermitian counterpart $H_{S}=\eta\mathcal{H}_{S}\eta^{-1}$
is given by

\begin{align}
H_{S} & =\eta\mathcal{H}_{S}\eta^{-1}=\frac{1}{2}\left(\omega-1-\omega^{2}\epsilon^{2}\right)p^{2}+\frac{1}{2}\frac{\omega^{2}-4\omega^{2}\epsilon^{2}}{\omega-1-\omega^{2}\epsilon^{2}}x^{2},
\end{align}
or in terms of the bosonic operators $b\left(b^{\dagger}\right)$,
$H=\Omega b^{\dagger}b,$ where we have discarded any zero point energy.
For $\epsilon\in\left(0,1/2\right)$, the Hamiltonian $\mathcal{H}_{S}$
is in the $\mathcal{PT}$- symmetric phase, with $\epsilon_{cr}=1/2$
usually representing the exceptional point of $\mathcal{H}_{S}$,
while for $\epsilon\in\left(1/2,\infty\right)$, $\mathcal{H}_{S}$
is in the $\mathcal{PT}$- symmetry broken phase.

Following the scheme of the previous section, the thermal state in
the standard quantum mechanics is 

\begin{equation}
\rho^{th}\left(\vec{\theta}\right)=\frac{\exp\left[-\beta H_{S}\right]}{Z},\label{thermalstate}
\end{equation}
where $Z=\sum_{n}\exp\left[-\beta H_{S}\right]$ is the partition
function, and $\vec{\theta}=\left(\omega,T,\epsilon\right)$ is a
vector of all parameters characterizing the state. All the relevant
physical effect concerning the  n-HC is in $\epsilon$.
For completeness, the covariance matrix using the vector of quadrature
operators for the state in Eq. (\ref{thermalstate}) is $\boldsymbol{\sigma}=\coth\left(\frac{\hbar\Omega}{2T}\right)\mathbb{I}_{2\times2}$,
with null first moments. We denote Eq. (\ref{thermalstate}) as the
probe state. There are two parameters that can be estimated from the
probe, the frequency $\omega$ and the temperature $T$. The frequency estimation
of a quantum harmonic oscillator is of particular interest and has
received considerable attention in the last years \citep{Garbe2020,Binder2020,Montenegro2025},
while the estimation of $T$ corresponds to quantum thermometry \citep{Cenni2022,Alves2024,Ullah2023}.

From Eq. (\ref{QFI}) we directly obtain $\mathcal{I}_{\omega}$ and
$\mathcal{I}_{\beta}$ as

\begin{align}
\mathcal{I}_{\omega}\left[\rho^{th}\left(\vec{\theta}\right)\right] & =\frac{1-4\epsilon^{2}}{4T^{2}\sinh^{2}\left(\frac{\hbar\Omega}{2T}\right)},\label{A}\\
\mathcal{I}_{T}\left[\rho^{th}\left(\vec{\theta}\right)\right] & =\frac{\omega^{2}\left(1-4\epsilon^{2}\right)}{4T^{2}\sinh^{2}\left(\frac{\hbar\Omega}{2T}\right)}.\label{B}
\end{align}

Equations (\ref{A}) and (\ref{B}) show explicitly the role played
by the  n-HC in the estimation of $\omega$ and
$T$. To compare fairly the gain ratio defined in Eq. (\ref{raito})
we note that Eq. (\ref{Swanson02}) becomes Hermitian if we set $\epsilon_{\text{Herm}}=\alpha/\omega$,
with $\alpha=1$, which results in $\eta=\mathbb{I}$ for the present
case, and 

\begin{equation}
\mathcal{H}=\mathcal{H}^{\dagger}=\omega a^{\dagger}a+a^{2}+a^{\dagger2}.
\end{equation}

\begin{figure}
\includegraphics[scale=0.46]{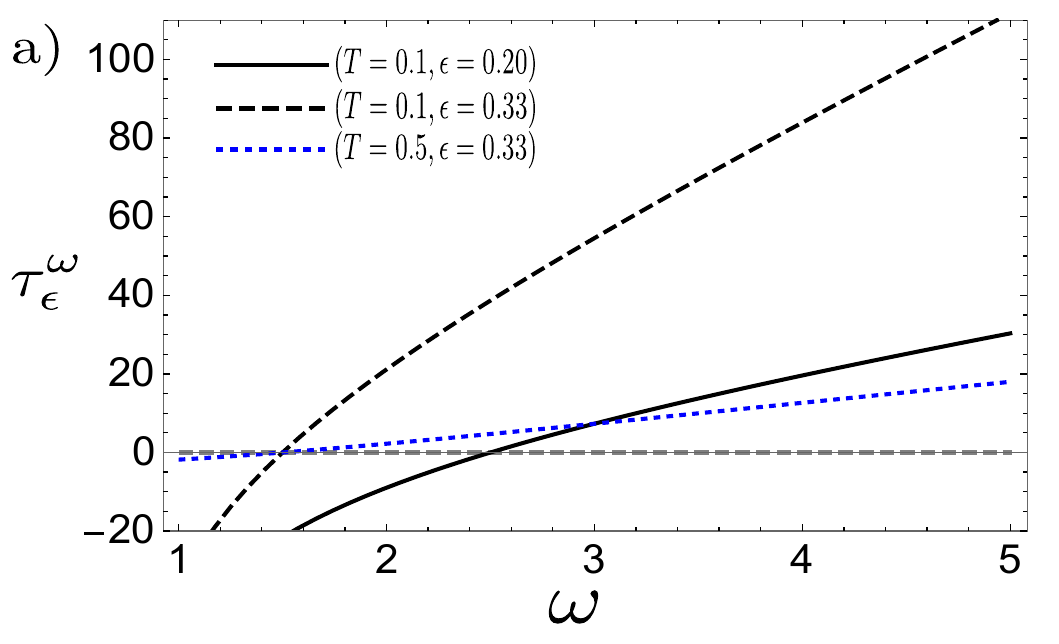}

\includegraphics[scale=0.5]{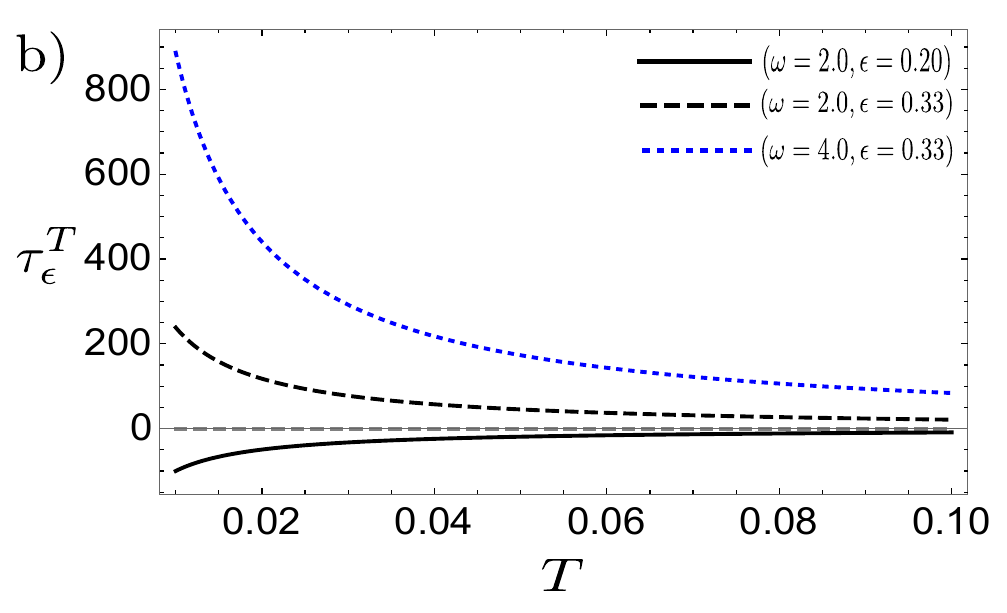}

\caption{Gain Ratio $\tau_{\epsilon}^{\omega}$ (a) and $\tau_{\epsilon}^{T}$
(b) as a function of the frequency $\omega$ and temperature $T$,
respectively. For $\tau_{\epsilon}^{\omega}$: $\left(T,\epsilon\right)=\left(0.1,0.2\right)$
(solid black line), $\left(T,\epsilon\right)=\left(0.1,0.3\right)$(dashed
black line), and $\left(T,\bar{\epsilon}\right)=\left(0.5,0.3\right)$
(dotted blue line). For $\tau_{\epsilon}^{\beta}$: $\left(\omega,\epsilon\right)=\left(2,0.2\right)$
(solid black line), $\left(\omega,\epsilon\right)=\left(2,0.3\right)$
(dashed black line), and $\left(\omega,\epsilon\right)=\left(4,0.3\right)$
(solid black line).}

\label{tau}
\end{figure}

Figure \ref{tau} depicts the gain ratio $\tau_{\epsilon}^{\omega}$
and $\tau_{\epsilon}^{T}$ as a function of $\omega$ and $T$, respectively.
For the frequency estimation (Figure \ref{tau}-(a)), by keeping $T$
fixed, as $\epsilon$ approaches the critical value $\epsilon_{cr}=0.5$,
see solid black and dashed black lines, the  n-HC
enhances the QFI for $\omega$ when compared with the case for $\epsilon_{\text{Herm}}$.
However, increasing the temperature degrades this enhancement, as
noted by the black dashed and blue dotted lines, a consequence of
fluctuation increasing in the average thermal number. For temperature
estimation (Figure \ref{tau}-(b)), we observe that for a fixed frequency,
the  n-HC does not represent an advantage if
$\epsilon$ is sufficient far from the critical value $\epsilon_{cr}$,
as seen by the solid black and dashed black lines. At the same time,
the blue dotted line shows that approaching the critical value $\epsilon_{cr}$
represents a considerable contribution in the QFI for $T$. Moreover,
any advantage for the temperature estimation decreases as $T$ increases,
for the same reason that for the frequency estimation and in an accordance with Ref. \cite{Serafinibook}.

The non-Hermitian contribution $H_{\text{NH}}=\alpha a^{2}+\beta a^{\dagger2}$
in the Swanson Hamiltonian as a feature to enhance the frequency and
temperature estimation can also be quantified in terms of a energetic
cost function \citep{Liuzzo-Scorpo2018,Lipka-Bartosik2018} to include the 
$H_{\text{NH}}$ in the Hamiltonian of the system. In fact, we can
determine the QFI per unit of the energetic cost during the estimation
of $\omega$ or $T$. We introduce the quantity 

\begin{equation}
u^{\theta_{i}}=\frac{\mathcal{I}_{\theta_{i}}\left[\rho^{th}\left(\vec{\theta}\right)\right]}{\Delta U},\label{uu}
\end{equation}
with $\Delta U=U\left[\rho^{th}\left(\vec{\theta}\right)\right]-U\left[\rho_{HO}^{th}\left(\vec{\theta}\right)\right]$,
where $U\left(\rho\right)=Tr\left[H\rho\right]$, with $H$ the respective
Hamiltonian, $\rho^{th}\left(\vec{\theta}\right)$ and $\rho_{HO}^{th}\left(\vec{\theta}\right)$
probe states based on the Hamiltonian in Eq. (\ref{Swanson02}) with
and without ($\alpha=0$) the $H_{\text{NH}}$ term. The explicit
form for $\Delta U$ in this case is

\begin{equation}
\Delta U=2\omega\left[\coth\left(\frac{\Omega}{2T}\right)-\coth\left(\frac{\omega}{2T}\right)\right].
\end{equation}

\begin{figure}
\includegraphics[scale=0.42]{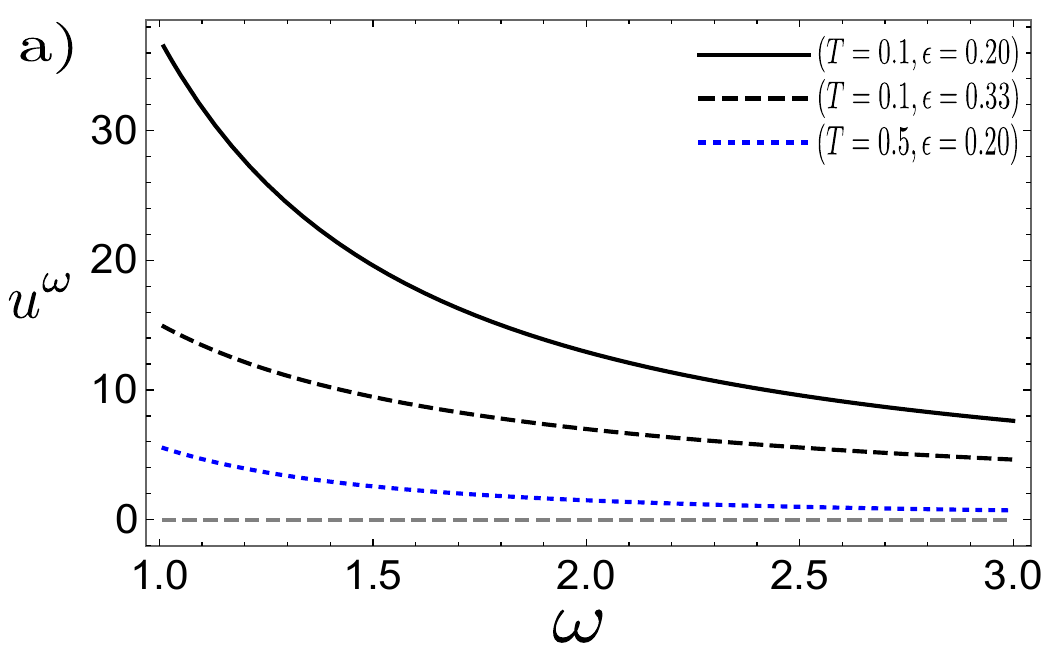}

\includegraphics[scale=0.46]{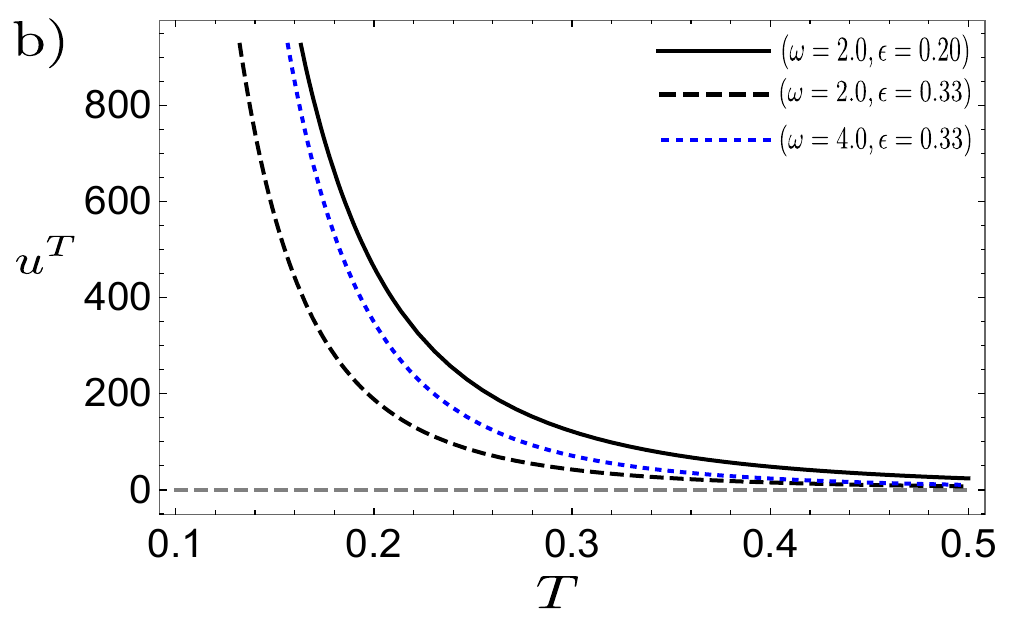}

\caption{Ratio between the quantum Fisher information and the energetic cost,
$u^{\theta_{i}}$, for the frequency and temperature estimation. a)
$u^{\omega}$ as a function of $\omega$ for $\left(T,\epsilon\right)=\left(0.1,0.2\right)$(black
solid line), $\left(T,\epsilon\right)=\left(0.1,0.33\right)$(dashed
solid line), and $\left(T,\epsilon\right)=\left(0.5,0.2\right)$ (blue
dotted line). b) $u^{T}$ as a function of $T$ for $\left(\omega,\epsilon\right)=\left(2,0.2\right)$(black
solid line), $\left(\omega,\epsilon\right)=\left(2,0.33\right)$(dashed
solid line), and $\left(\omega,\epsilon\right)=\left(4,0.33\right)$(blue
dotted line).}

\label{energy}
\end{figure}

Figure \ref{energy} illustrates the ratio between the quantum Fisher
information and the energetic cost for the frequency estimation $u^{\omega}$
as a function of $\omega$ and for the temperature estimation $u^{T}$
as a function of $T$. In Fig. \ref{energy}-a) we observe the for any
value of $\epsilon<\epsilon_{cr}$ the quantity $u^{\omega}$ decreases
as a function of $\omega$, indicating that for larger frequencies
the energy difference makes the choice of $\epsilon$ irrelevant.
Furthermore, we note that for the chosen temperature values, $u^{\omega}$
is higher for $\epsilon=0.2$ than $\epsilon=0.33$ (solid and dashed
black lines). On the other hand, the dotted blue line shows that increasing
the temperature decreases not only the QFI but also the ratio $u^{\omega}$.
Figure \ref{energy}-b) depicts the same but for the temperature estimation.
As the temperature is increased the quantity $u^{T}$ is degraded
irrespective the value of $\epsilon$. This is mainly a consequence
of the larger energetic cost as $T$ increases, which suppresses any
advantage due to $\epsilon$.

To conclude this part, we would like to highlight that if we had assumed the parameters $\alpha$  and $\beta$ regarding the n-HC to be $\alpha\beta \propto - \epsilon^2$, the enhancement in the QFI would not be achievable. We also considered this case in our simulations. The reason is that this assumption prevents the closing of the energy gap during the frequency and temperature estimation.

\subsection*{Estimation of the similarity transformation}

We may wonder about the estimation of the similarity transformation
given an ensemble of probe states. Again, for simplicity, we assume
the probe states to be given by Eq. (\ref{thermalstate}). Using Eq.
(\ref{QFI}) to compute the QFI for $\epsilon$, we have

\begin{equation}
\mathcal{I}_{\epsilon}\left[\rho^{th}\left(\vec{\theta}\right)\right]=\frac{4\epsilon^{2}\omega^{2}}{T^{2}\left(1-4\epsilon^{2}\right)\sinh^{2}\left(\frac{\hbar\Omega}{2T}\right)}.\label{Imu}
\end{equation}

\begin{figure}
\includegraphics[scale=0.4]{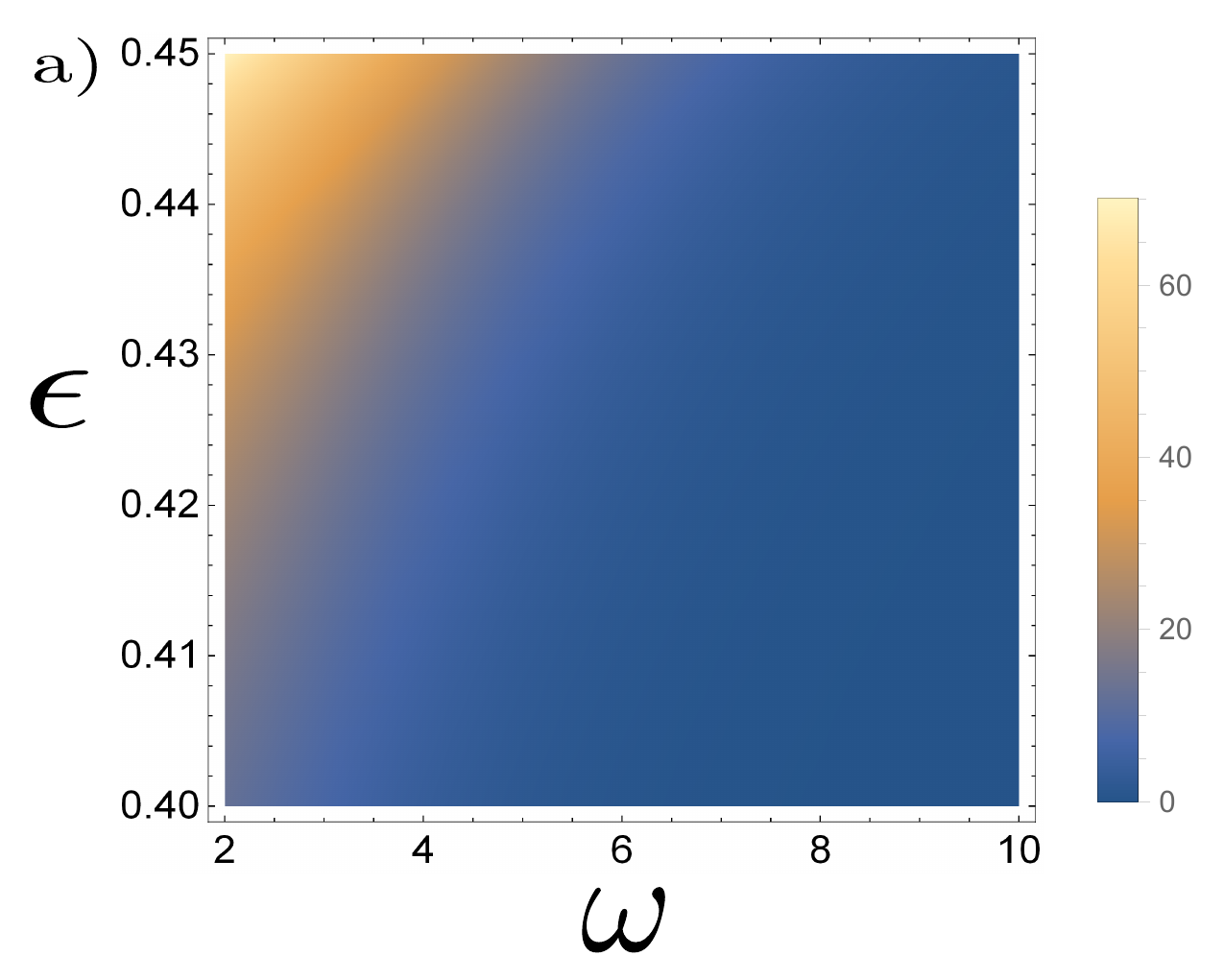}

\includegraphics[scale=0.41]{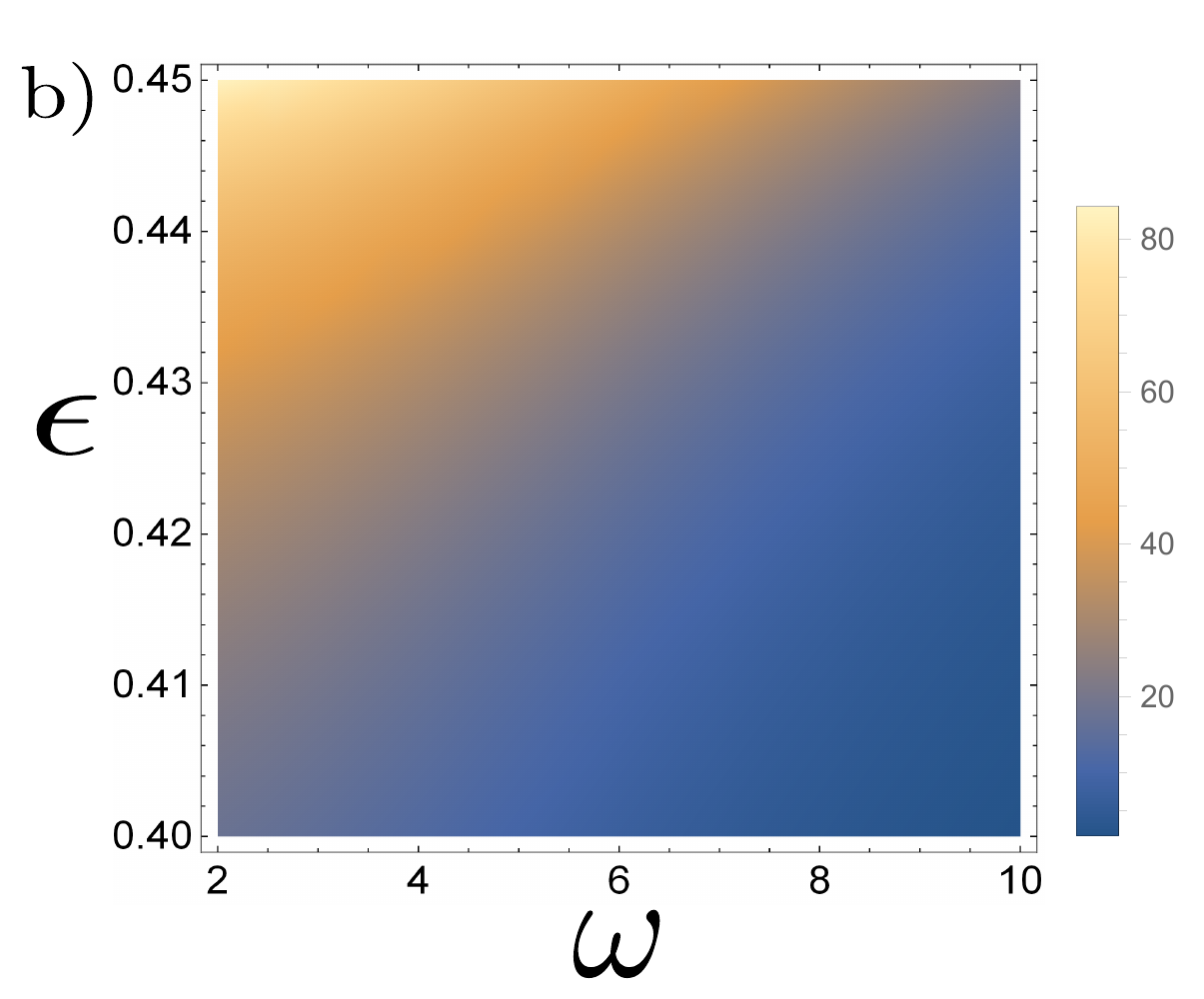}

\caption{Quantum Fisher information $\mathcal{I}_{\epsilon}\left[\rho^{th}\left(\vec{\theta}\right)\right]$
for the estimation of the similarity transformation parameter as function
of $\epsilon$ and $\omega$. Figures (a) and (b) are for $T=0.5$
and $\omega=1.0$, respectively.}

\label{QFImu}
\end{figure}

Figure \ref{QFImu} illustrates $\mathcal{I}_{\epsilon}$ as a density
plot as a function of $\epsilon$ and $\omega$, for $T=0.5$ and
$T=1.0$, Fig. \ref{QFImu}-a) and \ref{QFImu}-b), respectively.
Firstly, we observe that the higher the temperature, the higher the
maximum $\mathcal{I}_{\epsilon}$ and the area in which $\mathcal{I}_{\epsilon}$
is positive. Furthermore, for a fixed value of $\epsilon$, the increase
of the frequency corresponds to a decreasing of $\mathcal{I}_{\epsilon}$,
while it increase as $\epsilon$ approaches the critical value. These
results indicates that the uncertainty in estimating the similarity
transformation can be reduced for small values of frequency and relative
high temperature, i.e., in the limit $\beta\omega\ll1$.

\section{Conclusion\label{sec:Conclusion}}

In this work, we have addressed the quantum parameter estimation problem
in the scope of $\mathcal{PT}$-symmetric quantum mechanics. The first
result is the extension of the relation between Hermitian and non-Hermitian
expectation values for mixed states, in particular thermal states.
This shows that, starting from a $\mathcal{PT}$-symmetric Hamiltonian,
it is possible to apply the similarity transformation to obtain its
Hermitian counterpart and then compute the quantum Fisher information
using the standard form in Eq. (\ref{QFI}) for Gaussian states.

The gain ratio, which computes the advantage in the QFI due to the
non-Hermitian contribution, showed that for the frequency estimation,
the QFI increases as $\epsilon$ approaches its critical value, whereas
increasing the temperature can degrade considerably this effect. Furthermore,
estimating the temperature results in a decreasing of the gain ratio
as a function of the temperature, an effect due to the larger average
thermal number. In contrast, for small temperatures to be estimated
the non-Hermitian contribution can enhance the QFI. We also computed
the ratio between the QFI and the energetic cost associated with the
inclusion of a non-Hermitian term in the Hamiltonian, showing that
this ratio is advantageous for the estimation of small frequencies
and temperatures in general, but it decreases as $\omega$ and $T$
become larger. In addition, we also computed the QFI for the one-parameter
family characterizing the similarity transformation, with the QFI
being proportional to the temperature and decreasing with the frequency,
while it diverges as $\epsilon$ approaches the critical value.

Our results are general in the sense that the only restriction was
to impose that $\alpha\beta\propto\epsilon$. It is important to stress
that for $\alpha\beta\propto-\epsilon$ there is no the closing of
the energy gap, which in turn will not result in a advantage in the
frequency and temperature estimation. We hope that the results presented
here could contribute in quantum metrology exploiting non-Hermitian
physics.
\begin{acknowledgments}
Jonas F. G. Santos acknowledges CNPq Grant No. 420549/2023-4, Fundect
Grant No. 83/026.973/2024, and Universidade Federal da Grande Dourados
for support.
\end{acknowledgments}

\section*{Appendix A: details about the density matrix and mean values\label{sec:Appendix:-details-about}}

Starting from Eq (\ref{eq:aa}) and employing $|\phi_{n}\left(t\right)\rangle=\eta\psi_{n}\left(t\right)\rangle$,
we have

\begin{align}
\rho^{th} & =\sum_{n}c_{n}|\phi_{n}\rangle\langle\phi_{n}|\\
 & =\sum_{n}c_{n}\left(\eta\psi_{n}\left(t\right)\rangle\right)\left(\langle\psi_{n}|\eta^{\dagger}\right)\\
 & =\eta\tilde{\rho}^{th}\eta^{\dagger}.
\end{align}

Considering now expectation values evaluated by thermal states, we
have that

\begin{align*}
\langle O\rangle_{\rho^{th}} & =Tr\left[O\rho^{th}\right]\\
 & =\sum_{n}\langle\phi_{n}\left(t\right)|O\rho^{th}|\phi_{n}\left(t\right)\rangle\\
 & =\sum_{n}\left(\langle\psi_{n}\left(t\right)|\eta^{\dagger}\right)O\rho^{th}\left(\eta|\psi_{n}\left(t\right)\rangle\right)\\
 & =\sum_{n}\langle\psi_{n}\left(t\right)|\eta^{\dagger}\left(\eta\mathcal{O}\eta^{-1}\right)\left(\eta\tilde{\rho}^{th}\eta^{\dagger}\right)\eta|\psi_{n}\left(t\right)\rangle\\
 & =\sum_{n}\langle\psi_{n}\left(t\right)|\Theta^{2}\mathcal{O}|\psi_{n}\left(t\right)\rangle\\
 & =\langle\Theta^{2}\mathcal{O}\rangle_{\tilde{\rho}^{th}},
\end{align*}
where from the fourth to fifth line we have used the cyclic property
of the total trace and that $\eta^{-1}\eta=\mathbb{I}$.

\end{document}